\documentclass[ALICE,manyauthors]{cernphprep}

\usepackage[comma,square,numbers,sort&compress]{natbib}
\usepackage{hyperref}
\usepackage{lineno}
\usepackage{color}
\usepackage{bm}

\definecolor{dgreen}{cmyk}{1.,0.,1.,0.2}
\definecolor{orange}{cmyk}{0.,0.353,1.,0.}

\newcommand{\mean}[1]{{\langle{#1}\rangle}}

\def \be {\begin{equation}}
\def \ee {\end{equation}}
\def \deta {\Delta\eta}
\def \la {\langle}
\def \ra {\rangle}

\def \pbpb {Pb--Pb} 
\def \auau {Au--Au} 
\def \pt {p_{\rm T}} 
\def \snn {\sqrt{s_{\rm NN}}}
\def \dndeta {\langle\mathrm{d}N_{\mathrm{ch}}/\mathrm{d}\eta\rangle}

\begin{document}%

\begin{titlepage}
\PHyear{2015}
\PHnumber{316}      
\PHdate{07 December}  
%

\title{Charge-dependent flow and the search for the Chiral Magnetic
  Wave in \pbpb~collisions at $\bm{\snn}$ = 2.76 TeV}
\ShortTitle{Charge-dependent flow and the search for the Chiral
  Magnetic Wave} 

\Collaboration{ALICE Collaboration\thanks{See
    Appendix~\ref{app:collab} for the list of collaboration members}}
\ShortAuthor{ALICE Collaboration} 

\begin{abstract}
We report on measurements of a charge-dependent flow using a novel
three-particle correlator with ALICE in \pbpb~collisions at the LHC,
and discuss the implications for observation of local parity violation
and the Chiral Magnetic Wave (CMW) in heavy-ion collisions.
Charge-dependent flow is reported for different collision centralities
as a function of the event charge asymmetry.  While our results are in
qualitative agreement with expectations based on the CMW, the nonzero
signal observed in higher harmonics correlations indicates a possible
significant background contribution.  We also present results on a
differential correlator, where the flow of positive and negative
charges is reported as a function of the mean charge of the particles
and their pseudorapidity separation.  We argue that this differential
correlator is better suited to distinguish the differences in positive
and negative charges expected due to the CMW and the background
effects, such as local charge conservation coupled with strong radial
and anisotropic flow.
\end{abstract}
\end{titlepage}
\setcounter{page}{2}

%
%
%
%

\section{Introduction}

Parity (P) is a major symmetry of classical physics, being present in
rigid-body dynamics, classical electrodynamics, and gravity.  In the
development of quantum mechanics, parity conservation was assumed.  It
was not until the 1950s~\cite{Lee:1956qn} that the possibility of
parity violation was considered, and soon after it was definitively
demonstrated experimentally in nuclear
decays~\cite{Wu:1957my,Garwin:1957hc}.  In the modern picture, P- and
CP-violation in weak interactions are widely established
experimentally and well understood theoretically.  In strong
interactions there is very little or no global P-violation, as
determined by measurements of the neutron electric dipole moment
~\cite{Smith:1957ht,Baker:2006ts}.  However, there is no
first-principles reason why P- and CP-violation should not exist in
strong interactions.  P- and CP-violation as a general feature of
quantum field theories was first explored in the
1970s~\cite{Lee:1973iz,Lee:1974ma}, and a proposal to use heavy-ion
collisions as a tool for studying P- and CP-violation first appeared
as early as the 1980s~\cite{Morley:1983wr}.  Specific proposals for a
search for local P-violating effects in heavy-ion collisions appeared
in the last
decade~\cite{Kharzeev:1998kz,Kharzeev:1998kya,Kharzeev:2000na,Voloshin:2000xf,Kharzeev:2004ey,Voloshin:2004vk}.

Collisions of heavy nuclei at ultra-relativistic energies create a
hot, dense medium that appears to have partonic degrees of freedom and
evolve hydrodynamically.  In non-central collisions the initial
overlap region is non-isotropic, which, due to particle interactions,
leads to a momentum-space anisotropy of the produced particles.  This
anisotropy can be described using a Fourier expansion of the azimuthal
distribution of particles~\cite{Voloshin:1994mz}.  Non-central
collisions are also characterized by large orbital momentum and, what
is important to this study, very large magnetic fields.  Numerical
estimates~\cite{Skokov:2009qp,Zhong:2014cda} indicate that at LHC
energies the field strength can be as large as $B \approx m_{\pi}^2/e
\approx$~10\textsuperscript{14}~T.  In a vacuum, the magnetic field
induced by the spectators decays in time quadratically ($B \propto
t^{-2}$) and the lifetime of the magnetic field at LHC energies is
extremely short, decreasing 6 orders of magnitude over the course of
0.5 fm$/c$.  However the presence of electrical charges (such as the
quarks in the QGP) means there is finite electrical conductivity.  By
Lenz's law the change in magnetic field is opposed by the charge
carriers in the conductor, so the temporal decay of the magnetic field
is significantly slowed~\cite{Tuchin:2013ie,McLerran:2013hla}.

The Chiral Magnetic Effect (CME)~\cite{Fukushima:2008xe} is a process
of charge separation with respect to the reaction plane.  In the QCD
vacuum there can exist gluonic configurations with nonzero topological
charge, which can be an instanton or a sphaleron.  At high
temperatures, the sphaleron rate is expected to be dominant.  The
presence of such gluonic configurations with topological charge is
what drives the P-violation process.  For example, in a region with
negative topological charge, left-handed quarks will become
right-handed, and right-handed quarks will remain right-handed.  The
strong magnetic field created in heavy-ion collisions interacts with
the magnetic moment of the quarks and orients the spins of of quarks
with positive (negative) electric charge to be parallel
(anti-parallel) to the field direction.  Under the assumption of
massless quarks, right-handed quarks have their spins and momenta
aligned.  This will cause positive (negative) quarks to move parallel
(anti-parallel) to the magnetic field, leading to a positive electric
current and thus a positive electric charge dipole.  Due to the chiral
symmetry restoration, $u$ and $d$ quarks have only their bare Higgs
mass, which is of the order of a few MeV$/c^2$.  This is sufficiently
small to regard the quarks as effectively massless.  Based on simple
geometrical arguments, the magnetic field direction is always normal
to the reaction plane, and therefore straightforwardly accessible to
experiment.  The Chiral Separation Effect (CSE)~\cite{Son:2009tf} is a
similar effect in which the presence of a vector charge, e.g. electric
charge, causes a separation of chiralities.  For example, the presence
of a net positive electric charge will induce a positive axial current
along the direction of the magnetic field, i.e. right (left) handed
quarks moving parallel (anti-parallel) to the magnetic field.

The CME can be summarized in a relatively simple equation:
\begin{equation}
\vec{J}_V = \frac{N_c e}{2\pi^2}\mu_A \vec{B},
\end{equation}
where $\vec{J}_V$ is the vector current (electric charge current in
this case), $N_c$ is the number of colors (3 in QCD but other numbers
of colors, i.e. $SU(N_c)$ gauge fields, are of interest in theory),
$\mu_A$ is the axial chemical potential (which encodes the
anomaly-induced chiral imbalance), and $\vec{B}$ is the magnetic
field.
The CSE can similarly be summarized as:
\begin{equation}
\vec{J}_A = \frac{N_c e}{2\pi^2}\mu_V \vec{B},
\end{equation}
where $\vec{J}_A$ is the axial current (flow of axial charges, i.e. chiralities) and
$\mu_V$ is the vector (electric) chemical potential.
The coupling between these two phenomena leads to a wave propagation
of the electric charge, resulting in an electric charge quadrupole
moment of the system.  This is called the Chiral Magnetic Wave
(CMW)~\cite{Kharzeev:2010gd,Burnier:2011bf,Burnier:2012ae}.
Importantly, for a given (net) charge state of the system, the
quadrupole moment always has the same sign and is therefore present in
an average over events with the same vector charge state, meaning it
may lead to a signal strong enough to be observed directly in
experiment.

As mentioned above, the azimuthal distribution of particles can be
written as a Fourier expansion:
\begin{equation}
\frac{dN}{d\varphi} \propto 1 + \sum_{n} 2v_n \cos(n(\varphi-\psi_n)),
\end{equation}
where $\varphi$ is the azimuthal angle of the particle, $v_n$ is the
Fourier coefficient, and $\psi_n$ is the symmetry plane, which in
principle can be different for each harmonic number $n$.

Taking into account the well-known modulation of particle emission due
to elliptic flow parameterized with the Fourier coefficient $v_2$ (see
e.g.~\cite{Abelev:2014mda}), one can write the azimuthal distribution
of charged particles due to the CMW~\cite{Burnier:2012ae} as
\begin{equation}
\frac{dN^{\pm}}{d\varphi} = N^{\pm}[1+(2v_2 \mp r A)\cos(2(\varphi-\psi_2))],
\label{eq:cmwfourier}
\end{equation}
where the charge asymmetry $A=(N^+-N^-)/(N^++N^-)$ is determined in
some kinematic region (for example in the experimental acceptance),
and the parameter $r$ encodes the strength of the electric quadrupole
due to the CMW.  Bjorken flow~\cite{Bjorken:1982qr} relates the
pseudorapidity of a particle to its longitudinal production point.
Due to space-momentum correlations, the charge asymmetry $A$,
determined in the experimental acceptance, corresponds to the local
charge asymmetry in a certain region of the fireball.

ALICE measurements of the charge dependent correlations in search for
the CME have been published in~\cite{Abelev:2012pa}.  This paper
presents the ALICE results on the charge dependent elliptic flow as a
function of the event charge asymmetry. As the event charge asymmetry
$A$ strongly depends on the experimental acceptance and tracking
efficiency, we also present related results on a differential three
particle correlator that allows much more detailed study of the
underlying physics mechanisms. We also present the corresponding
measurements for higher harmonics flow that should be largely
insensitive to the CMW but sensitive to the possible background
effects.

\section{Analysis methodology}

\subsection{$v_n$ as a function of $A$ and integral three-point correlators}

In the theoretical work on the
CMW~\cite{Kharzeev:2010gd,Burnier:2011bf,Burnier:2012ae,Taghavi:2013ena,Yee:2013cya}
as well as the analysis published by STAR~\cite{Adamczyk:2015kwa} of
\auau~collisions at $\snn$ = 200 GeV, the observable has been the
charge-dependent flow coefficient $v_n^\pm$ as a function of the
charge asymmetry $A$.  Experimentally, the charge asymmetry defined in
a specified kinematic region must be corrected for detector
efficiency, as discussed
in~\cite{Adamczyk:2015kwa,Wang:2012qs,Ke:2012qb}.  The effect of the
correction is to increase the slope of positive or negative particle
$v_n^\pm$ vs. $A$.  This correction to the slope, though absolutely
  necessary, has the undesired feature of introducing additional
  sources of systematic uncertainty.

In addition to measuring $v_n^\pm$ as a function of $A$ for a
specified event-selection criterion (e.g. a centrality interval), one
can measure the covariance of $v_n^\pm$ and $A$, i.e. $\mean{v_n^\pm
  A} -\mean{A}\mean{v_n^\pm}$, as a function of some event-level
variable (e.g. centrality)~\cite{Voloshin:2014gja}.  The harmonic
coefficient is, by definition, $v_n
=\langle\cos(n(\varphi_1-\psi_n))\rangle$, where $\varphi_1$ is the
azimuthal angle of a particle in the event and $\psi_n$ is the $n$-th
harmonic symmetry plane.  This makes $\langle v_n^\pm A\rangle -
\langle A\rangle\langle v_n^\pm\rangle$ a three-point correlator.  The
first point is the flow particle, the second point is the event plane
(which is an estimator for the true symmetry plane), and the third
point is the event charge asymmetry.  In cases where the event plane
is determined with a second particle, as is the case with the
two-particle cumulant method, this correlator can also be called a
three-particle correlator.

The strength of the covariance of any two variables is independent of
the sample size as long as the correlation is statistically
significant, so no correction for efficiency is needed for the
three-point correlator.  For that reason, the correlation
strength is identical for the full set of particles in some collection
of events and for some randomly selected subset of particles in the
same collection of events~\cite{Belmont:2014lta}.

From Eq.~\ref{eq:cmwfourier}, it follows that $v_2^\pm \approx
\overline{v}_2 \mp rA/2$, which in turn allows one to write $\Delta
v_2=v_2^--v_2^+ \approx rA$.  By measuring $\Delta v_2$ as a function
of $A$, it is possible to extract $r$ directly in experiment.  One can
also substitute these terms into the three particle correlator, giving
\begin{equation}
\mean{v_2^\pm A} -\mean{A}\mean{v_2^\pm} \approx \mp
r\left(\mean{A^2}-\mean{A}^2\right)/2= \mp r\sigma^2_A /2.
\end{equation}
Clearly, either of these approaches can be generalized to arbitrary
harmonic $v_n$, yielding in principle a different $r$ for each
harmonic.  Because of the symmetry of the CMW effect,
the expectation is that $r$, and therefore the
three-particle correlator, is significant
 only for the second harmonic and strongly suppressed for higher
   harmonics.
Since $A$ is efficiency dependent, it is necessary to scale down the
observed width to account for the natural broadening due to the binomial
sampling.  To calculate the scale factor, we compared the widths of $A$
in a Monte Carlo simulation before and after reconstruction effects,
where the full detector response was implemented within the
GEANT3~\cite{Brun:1994aa} framework.  Since the ALICE tracking
efficiency is independent of centrality, this scale factor is also
independent of centrality.  This correction to $\sigma_A^2$
introduces an additional
source of systematic uncertainty, in essentially the same way as the
correction to the slope for $v_n^\pm$ vs $A$.  In the present
analysis, this uncertainty represents a roughly 6\% normalization
uncertainty on the points.  Both the correction and the associated
uncertainty vary depending on detector acceptance and efficiency,
analysis cuts, etc.  A key advantage of measuring the three-particle
correlator, rather than $r$ (directly or indirectly), is that it is
efficiency independent and does not require any correction.  This
reduces the overall systematic uncertainty on the measurement, which
better facilitates comparisons across experiments as well as with
theory calculations.

In this analysis, two-particle cumulants are always used to calculate
$v_n$ (which ignores correlations not related to anisotropic flow
(non-flow), as well as flow fluctuations).  Then, the integral
correlator is:
\begin{eqnarray}
\langle v_n^{\pm}A\rangle - \langle A\rangle \langle v_n\rangle
&=& \la \cos[n(\varphi_1-\psi_n)] \, A\ra -\la
\cos[n(\varphi_1-\psi_n)] \ra \la A\ra \nonumber \\
&\approx& \frac{\la
  \la \cos[n(\varphi_1-\varphi_2)] A\ra\ra } {\sqrt{\la \la
    \cos[n(\varphi_1-\varphi_2)] \ra \ra }} -\sqrt{\la \la
  \cos[n(\varphi_1-\varphi_2)] \ra \ra } \la A\ra.
\label{eq:corr}
\end{eqnarray}
For this equation the inner average represents an average over all
particles in a single event, and the outer average represents an
average over all events.  The first particle has the selected charge
and the second particle is of both charges.  For the integral
correlator reported below, the same particles are used to calculate
$v_n$ and $A$.  The slopes extracted from the three-particle
correlator were checked against the slopes extracted directly from
$\Delta v_2$ vs. $A$ and found to be perfectly consistent with each
other.

\subsection{Differential three-point correlators}

A key advantage of the novel three-point correlator is that it permits
more differential studies and as such has more discriminating power.
The charge asymmetry in the event can be generalized to the charge of
a particle in the event, which we will call $q_3$.  The average of all
charges in the event is equal to the charge asymmetry, i.e.  $\langle
q_3\rangle_{\text{event}} \equiv A$.  Under this generalization the
correlator (Eq.~\ref{eq:corr}) becomes $\langle v_nq_3\rangle -
\langle q_3\rangle\langle v_n\rangle$.

We also use the additional notation of $\langle q_3\rangle_1$ for
denoting the mean of $q_3$ evaluated when selecting on the charge of
the first particle $q_1$.  This is important because, by construction,
the correlator $\langle v_nq_3\rangle - \langle q_3\rangle \langle
v_n\rangle$ contains reducible correlations, i.e. correlations that
can be expressed in terms of lower order
correlations~\cite{Belmont:2014lta,Voloshin:2014gja}.  These reducible
correlations are removed by the construction $\langle v_nq_3\rangle -
\langle q_3\rangle_1 \langle v_n\rangle$, which is therefore a
three-point cumulant.

Using these relations, we estimate the differential correlator in the following way:
\begin{eqnarray}
\langle v_n^{\pm}q_3\rangle - \langle q_3\rangle_1 \langle v_n\rangle
&=& \la \cos[n(\varphi_1-\psi_n)] \, q_3\ra -\la
\cos[n(\varphi_1-\psi_n)] \ra \la q_3\ra_1 \nonumber \\
&\approx& \frac{\la
  \la \cos[n(\varphi_1-\varphi_2)] q_3\ra\ra } {\sqrt{\la \la
    \cos[n(\varphi_1-\varphi_2)] \ra \ra }} -\sqrt{\la \la
  \cos[n(\varphi_1-\varphi_2)] \ra \ra } \la q_3\ra_1.
\end{eqnarray}

The evaluation of a differential correlator is a very important
feature of this study.  Rather than measuring only event quantities,
one can also measure the relationship between the flow at a particular
kinematic coordinate and the charge of the third particle at another
particular coordinate.  This means the effect can be measured as a
function of the separation in pseudorapidity of particles 1 and 3, for
example.  This differential nature allows for a much more detailed
study of the origin of the correlation and provides stronger
experimental constraints on the theoretical modeling of such effects.

Throughout this paper we use the subscript notation used above for the
first, second, and third particles.  The charge, azimuthal angle, and
pseudorapidity of the first particle are $q_1$, $\varphi_1$, and
$\eta_1$, respectively.  Similarly, the azimuthal angle of the second
particle is $\varphi_2$, and the charge and pseudorapidity of the
third particle are $q_3$ and $\eta_3$, respectively.

\section{Experimental apparatus and data analysis}

ALICE~\cite{ALICE,Abelev:2014ffa} is a dedicated heavy-ion experiment
located at the Large Hadron Collider at CERN.  It is composed of a
wide array of detector subsystems.  Those used in the present analysis
are the V0 detectors, the Inner Tracking System (ITS), and the Time
Projection Chamber (TPC).
The V0 detectors consist of scintillator arrays and are used for
triggering and centrality determination.  There are two V0 detectors,
V0A and V0C.  The V0A is located 340 cm from the nominal interaction
point and the V0C is installed at 90 cm distance in the opposite
direction.  The V0A covers 2.8~$<\eta<$~5.1 in pseudorapidity and the
C-side spans -3.7~$<\eta<$~-1.7.
The ITS is used for both tracking and vertex determination.  The ITS
is composed of three subsystems, each having two cylindrical layers of
silicon detectors.  Each of the layers covers at least $|\eta|<$~0.9
in pseudorapidity to match the TPC acceptance.
The TPC is the primary tracking detector at midrapidity.  The TPC is a
large gas volume detector separated into two regions by a central
electrode, positioned in a solenoidal magnetic field of 0.5~T.  The
gas volume is contained in a cylindrical electric field cage with an
inner radius of 85~cm and an outer radius of 2.5 m, spanning the full
azimuth $0<\varphi<2\pi$.  It extends 5.0~m in the $z$-direction,
providing coverage of the full radial track length for pseudorapidity
$|\eta|<$~0.9.

The present manuscript reports an analysis of \pbpb~collisions at
$\snn$ = 2.76 TeV, collected by ALICE during the 2010 and 2011 years
of LHC operations.  In the early part of the 2010 operation, the
\pbpb~minimum bias (MB) trigger was a 2-out-of-3 coincidence of a)
signals in two pixel chips in the outer layer of the SPD, b) a signal
in the V0A, c) a signal in the V0C.  In the later part of the 2010
operation and for all of the 2011 operation, the \pbpb~MB trigger
required a coincidence of both V0 detectors.
The data sample used in this analysis comprises approximately
1.7$\times$10$^7$ MB triggered events in the 2010 data set.
In the 2011 set, we use a mix of the central, semi-central, and MB
triggers.  The central trigger is an online trigger with a threshold
on the multiplicity in the V0 detectors such that it corresponds to
the 10\% most central events.  The semi-central trigger is defined
similarly such that it corresponds to the 50\% most central events.
The centrality is estimated using the mean multiplicity in the V0
detectors, and the centrality is required to be within 5\% (absolute)
of the centrality estimate using the TPC multiplicity to avoid
multiplicity fluctuations in the central region.  The longitudinal
position of the primary vertex is required to be within 10~cm of the
nominal center of the ALICE coordinate system in order to ensure
uniform detector acceptance.

Tracks are selected in the kinematic region $|\eta|<$~0.8 and
0.2~GeV/$c$~$<\pt<$~5.0~GeV/$c$.  They are required to have at least
70 TPC clusters, and the percentage of registered hits to crossed TPC
pad rows to be at least 80\%.  The track fit is required to have
$\chi^2$ per cluster (2 degrees of freedom) less than 4.0.  Additional
tracking information from the ITS is used when it is available,
i.e. when the track trajectory in the TPC points to an active area of
the ITS.  The distance of closest approach to the reconstructed vertex
is required to be within 3.2~cm in the $z$-direction and within 2.4~cm
in $xy$-plane.  Due to the excellent azimuthal uniformity of the TPC
response, no correction for azimuthal acceptance is needed, nor is one
applied.  The results are corrected for the $\pt$ dependence of the
tracking efficiency, which is about 80\% at low $\pt$ and about 90\%
at high $\pt$.  The correction procedure is to randomly exclude tracks
in such a way that the effective efficiency is made to be uniform in
$\pt$.  The result is a 2--3\% reduction in $v_n$.

To assess systematic uncertainties, the analysis is repeated for
different operational conditions (i.e. the two orientations of the
experimental magnetic field), different event selection criteria,
different track selection cuts, and different track reconstruction
methods.  The uncertainties associated with each specific selection or
condition are observed to be uncorrelated and thus added in quadrature
to determine the overall systematic uncertainty.  All aforementioned
sources were found to generally contribute with similar magnitude.
Many observables reported in this manuscript have values very close to
zero, so that reporting systematic uncertainties as a percentage of
these values obscures their true meaning.  For numerical stability,
systematic uncertainties are evaluated as a percentage of $\langle v_n
\rangle\langle A\rangle$ or $\langle v_n \rangle\langle q_3\rangle$.
This quantity alone is not necessarily physically meaningful, because
it contains detector specific effects.  It does, however, set a
natural scale for the uncertainties.  Once uncertainties are assessed,
their absolute value is determined and then plotted together with the
data points.

\section{Results}

\subsection{$v_2$ vs. $A$}

\begin{figure}[h!]
\begin{center}
\includegraphics[width=0.49\linewidth]{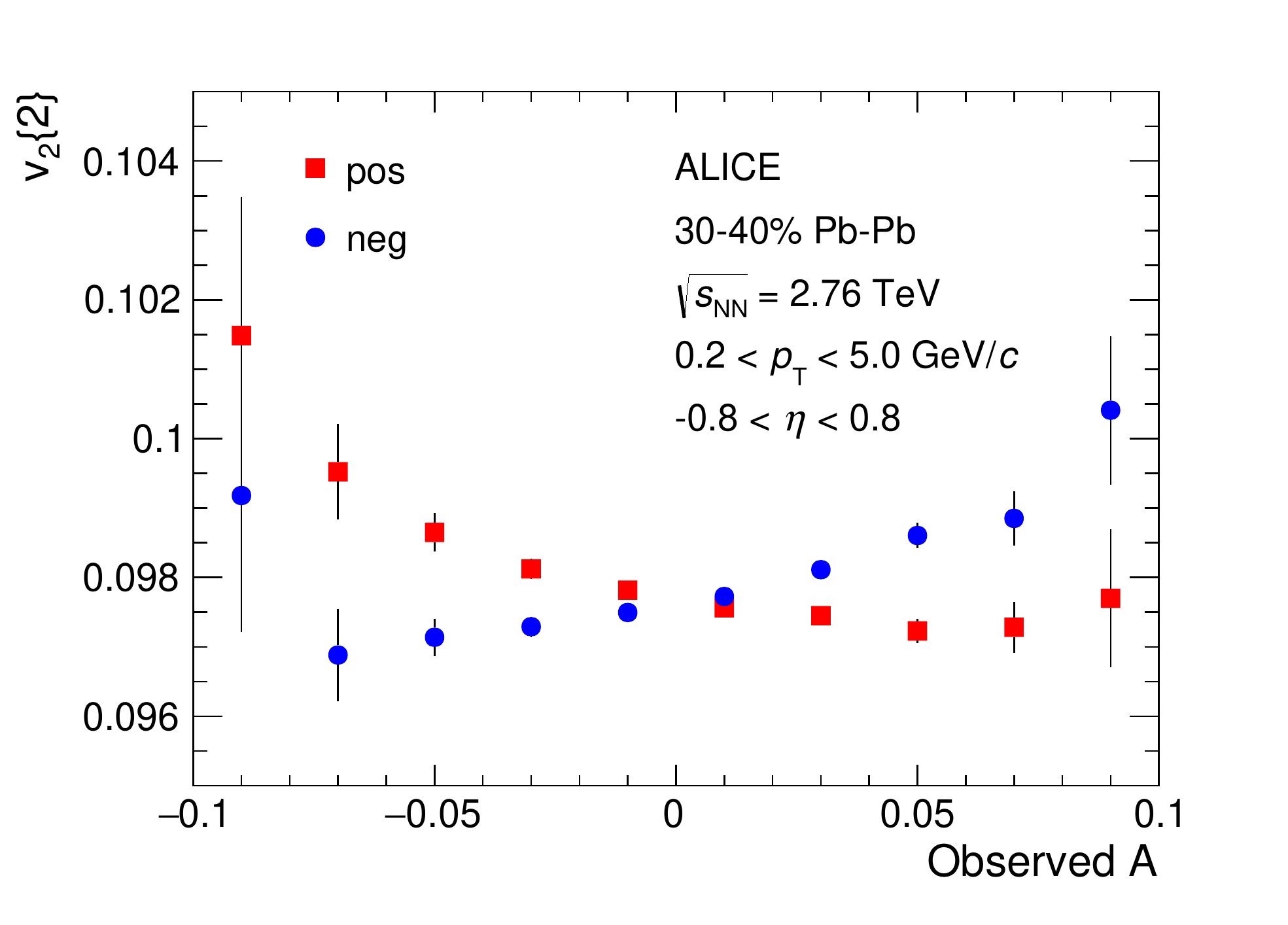}
\caption{(Color online) Harmonic coefficients $v_2^+$ (red squares)
  and $v_2^-$ (blue circles) as a function of the observed event
  charge asymmetry $A$ in the 30--40\% centrality class. Statistical
  uncertainties only.}
\label{fig:v2vsAcent34}
\end{center}
\end{figure}

Figure~\ref{fig:v2vsAcent34} shows $v_2^+$ and $v_2^-$ as a function
of the observed (uncorrected) event charge asymmetry $A$ in the
30--40\% centrality class.  Clearly visible is an increase in $v_2^-$
with increasing $A$, along with a corresponding decrease in $v_2^+$.
This is qualitatively consistent with expectations from the
CMW~\cite{Kharzeev:2010gd,Burnier:2011bf,Burnier:2012ae,Taghavi:2013ena,Yee:2013cya}
as well as with the STAR results~\cite{Adamczyk:2015kwa}.  Visually
the relationship does not appear exactly linear for either charge.
For a given centrality selection, there is some range of
multiplicities.  Since $A$ is a combination of numbers of particles,
it is a combination of different negative binomial distributions,
which are broader for lower numbers of particles.  Therefore, for
larger values of $|A|$, one is sampling events with lower
multiplicities, which can affect the value of
$v_2$.

\begin{figure}[h!]
\begin{center}
\includegraphics[width=0.49\linewidth]{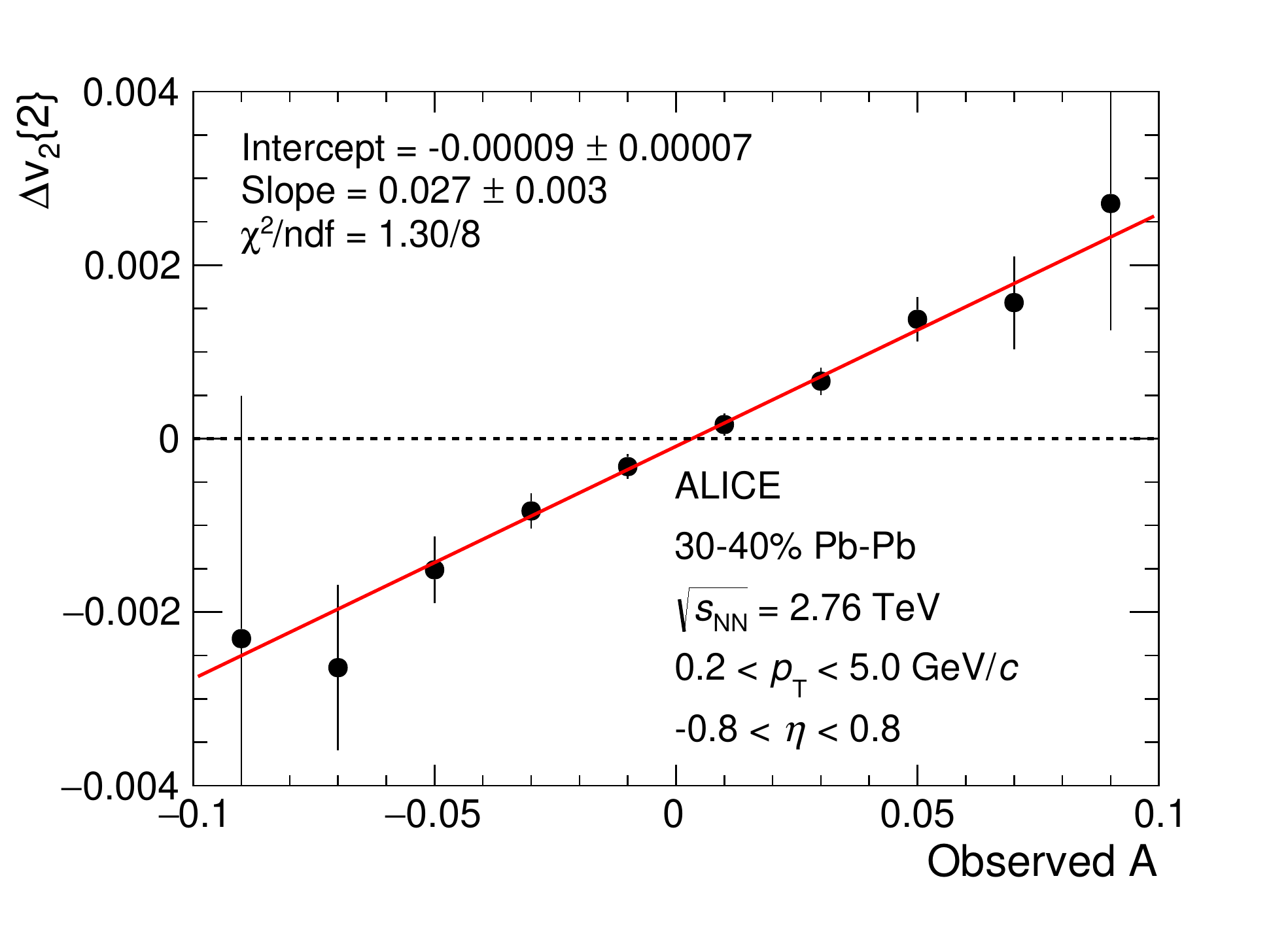}
\includegraphics[width=0.49\linewidth]{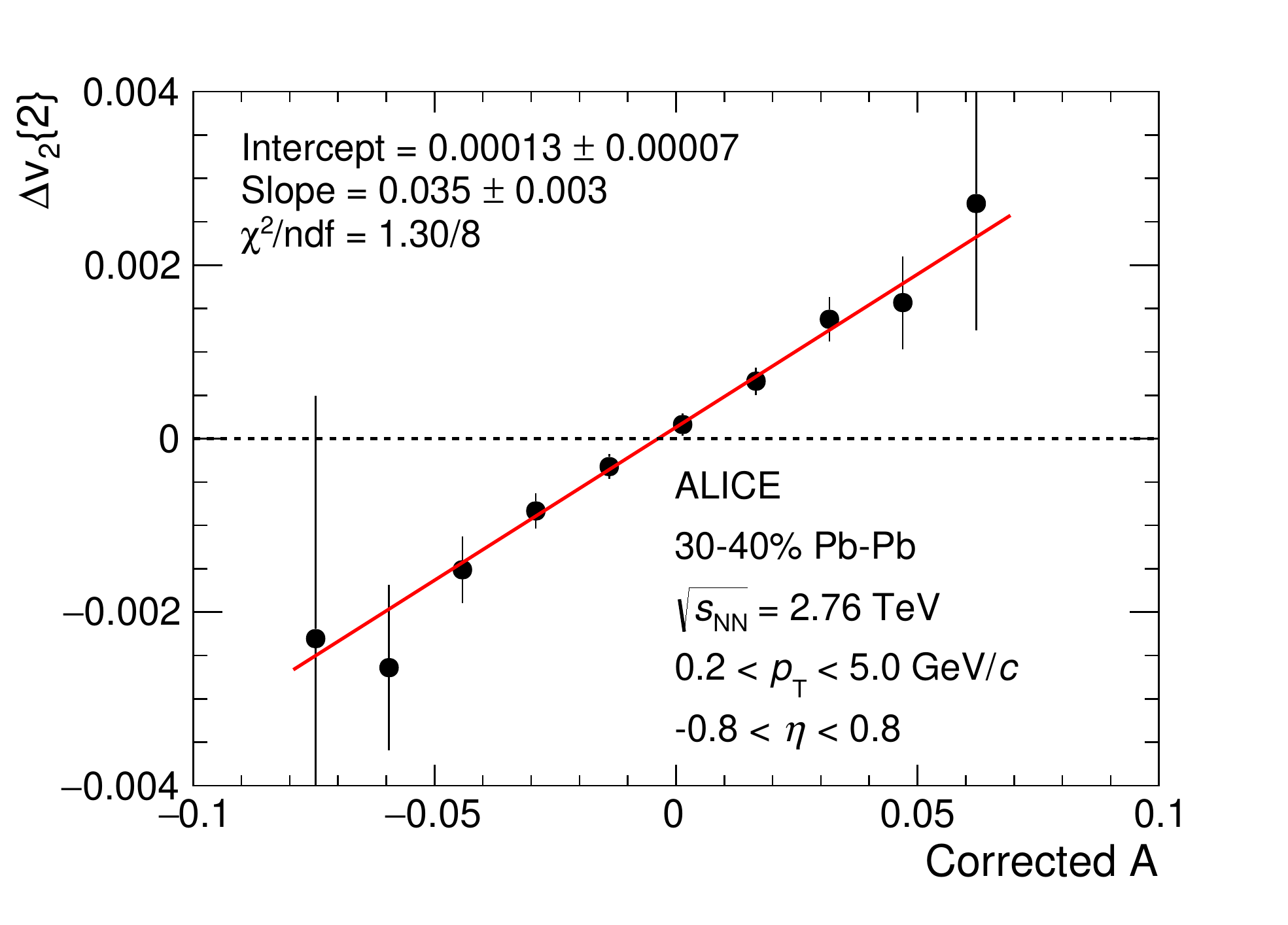}
\caption{(Color online) $\Delta v_2 = v_2^--v_2^+$ as a function of
  the observed (left) and corrected (right) event charge asymmetry $A$
  in the 30--40\% centrality class.  Statistical uncertainties only.}
\label{fig:Dv2vsAcent34}
\end{center}
\end{figure}

Figure~\ref{fig:Dv2vsAcent34} shows $\Delta v_2 = v_2^--v_2^+$ as a
function of the observed $A$ in the left panel and of $A$ corrected
for efficiency in the right panel.  To obtain the corrected $A$, we
analyzed HIJING~\cite{Wang:1991hta} simulations propagated through a
detector description implemented in the GEANT3~\cite{Brun:1994aa}
framework to determine the true (generated particle level) $A$ as a
function of the observed (reconstructed track level) $A$.  It can be
seen that the effect of the correction is a modest increase in the
slope.  Again, these results are qualitatively consistent with CMW
expectations and with the STAR data~\cite{Adamczyk:2015kwa}.

\subsection{Integral correlator results as a function of centrality}

Considering the observed increase (decrease) of $v_2^-$ ($v_2^+$) with
increasing $A$, discussed in the preceding section, we expect a
positive (negative) covariance of $v_2^-$ ($v_2^+$) with $A$, and
indeed this is exactly what is seen in the integral correlator.
Additionally, it enables a convenient study of the evolution of the
correlation as a function of event-level observables.

\begin{figure}[h!]
\begin{center}
\includegraphics[width=0.49\linewidth]{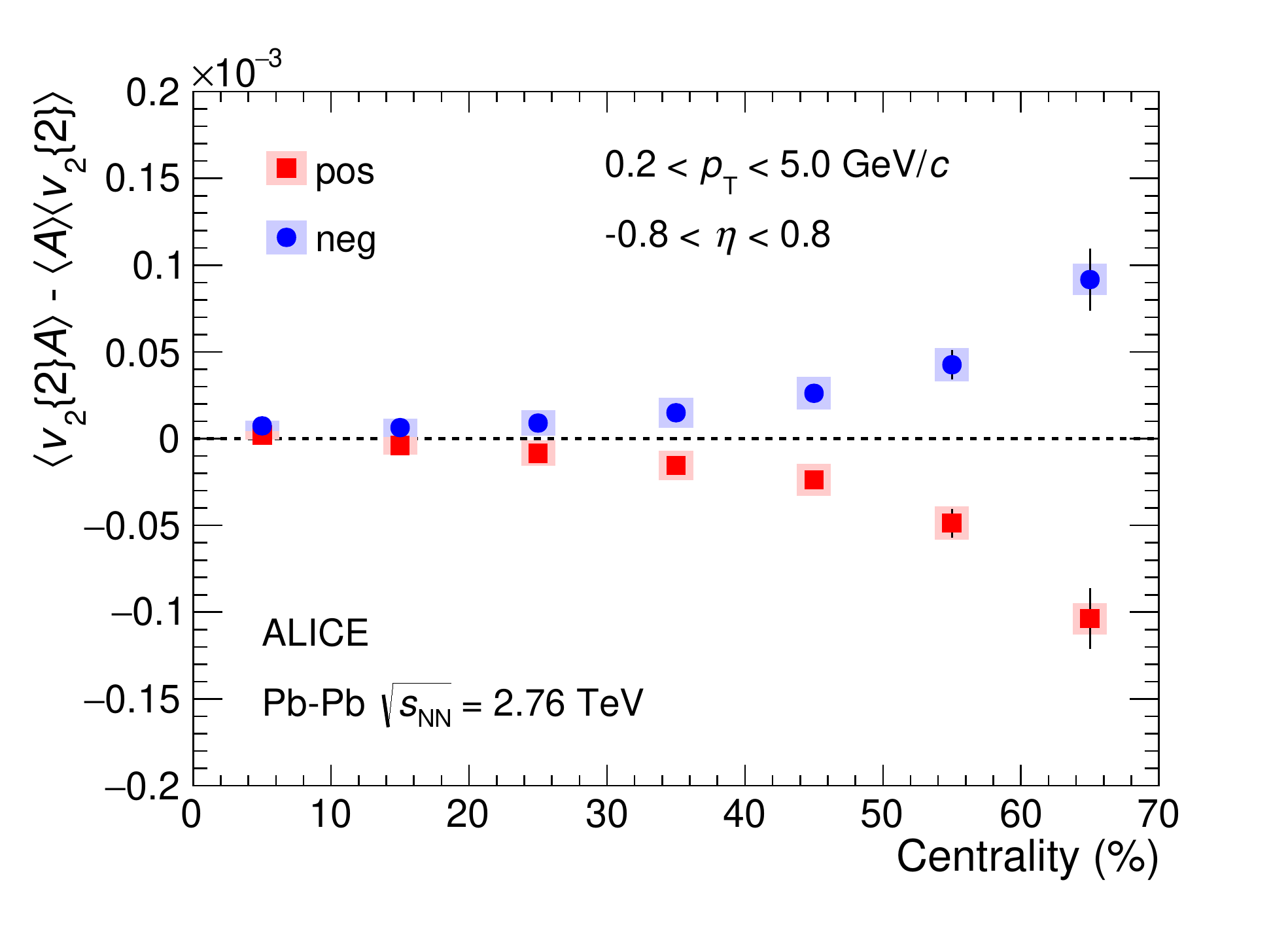}
\caption{(Color online) Three-particle correlator for positive (red
  squares) and negative (blue circles) particles for the second
  harmonic as a function of centrality.  Statistical (systematic)
  uncertainties are indicated by vertical bars (shaded boxes).}
\label{fig:c22Avscent}
\end{center}
\end{figure}

Figure~\ref{fig:c22Avscent} shows the integral correlator of the
second harmonic as a function of centrality.  A substantial increase
in the correlation strength is seen as the collisions become more
peripheral.  This can be caused by a combination of several factors.
The magnetic field strength increases as the impact parameter
increases since there are more spectators and thus the current gets
stronger.  This would cause the correlations due to the CMW to get
stronger.  Additionally, local charge conservation (LCC) effects could
play a role~\cite{Voloshin:2014gja}.  It is important to note that
neither of these necessarily comes at the expense of the other; in
principle the observable could have contributions from both of these
and/or additional contributions from as yet unknown sources of
correlation.

\begin{figure}[h!]
\begin{center}
\includegraphics[width=0.49\linewidth]{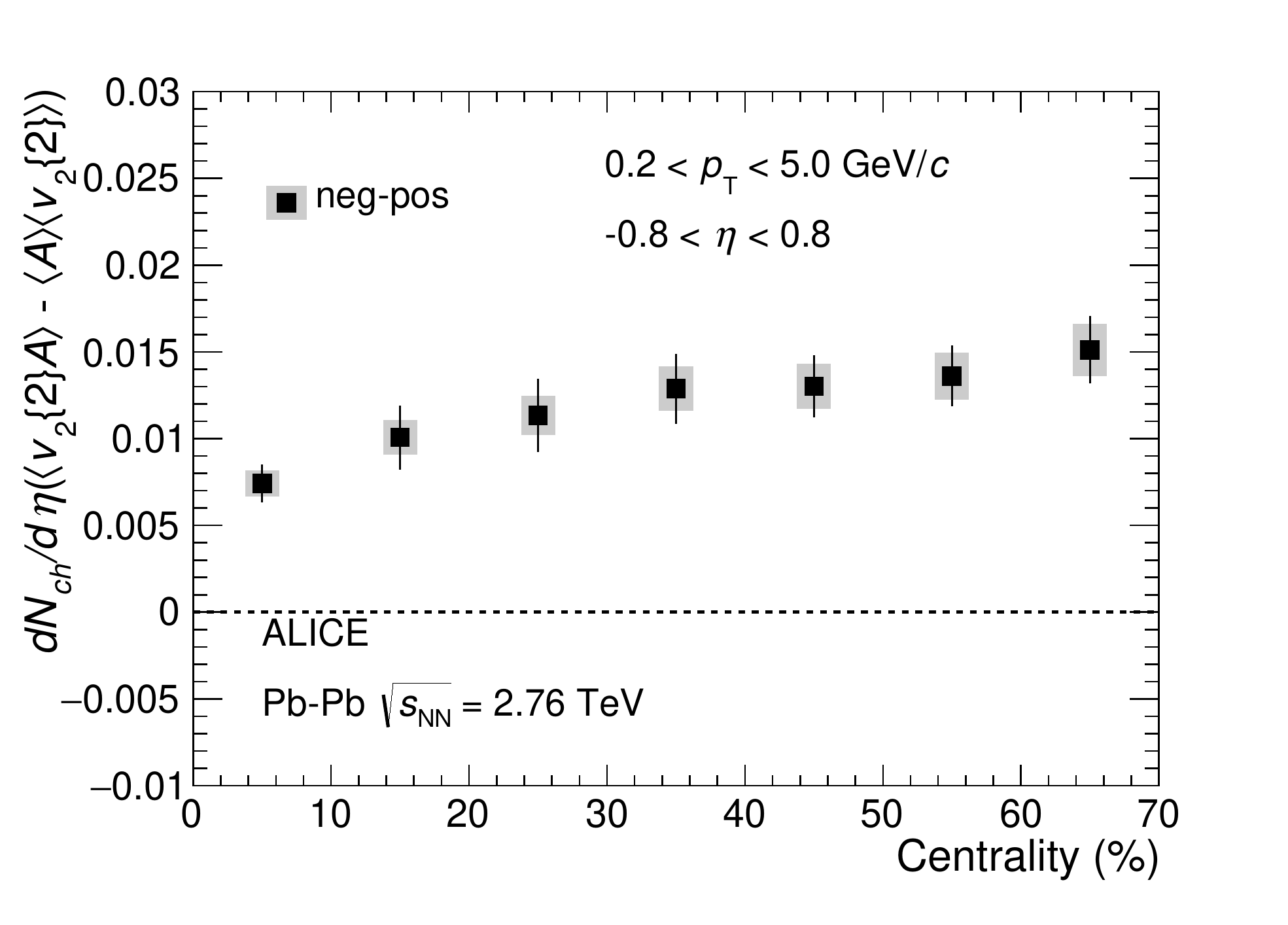}
\caption{Charge difference of the three-particle correlator for the
  second harmonic as a function of centrality, multiplied by
  $\dndeta$~\cite{Aamodt:2010cz}.  Statistical (systematic)
  uncertainties are indicated by vertical bars (shaded boxes).}
\label{fig:c22Adnvscent}
\end{center}
\end{figure}

Local charge conservation is the production of charged pairs at the
same spacetime point.  For $N$ particles, there are $N/2$ correlated
pairs, and $N(N-1)$ combinatoric pairs, meaning the correlation
strength is proportional to $N/(N(N-1))$, or approximately $1/N$.
Figure~\ref{fig:c22Adnvscent} shows the
difference between the charges for the second harmonic correlator
multiplied by $\dndeta$~\cite{Aamodt:2010cz}, where $\dndeta$ is used
as a proxy for the total number of particles, to examine the role
of the dilution of LCC correlations on the correlator.  Considerable
centrality dependence remains.

\begin{figure}[h!]
\begin{center}
\includegraphics[width=0.49\linewidth]{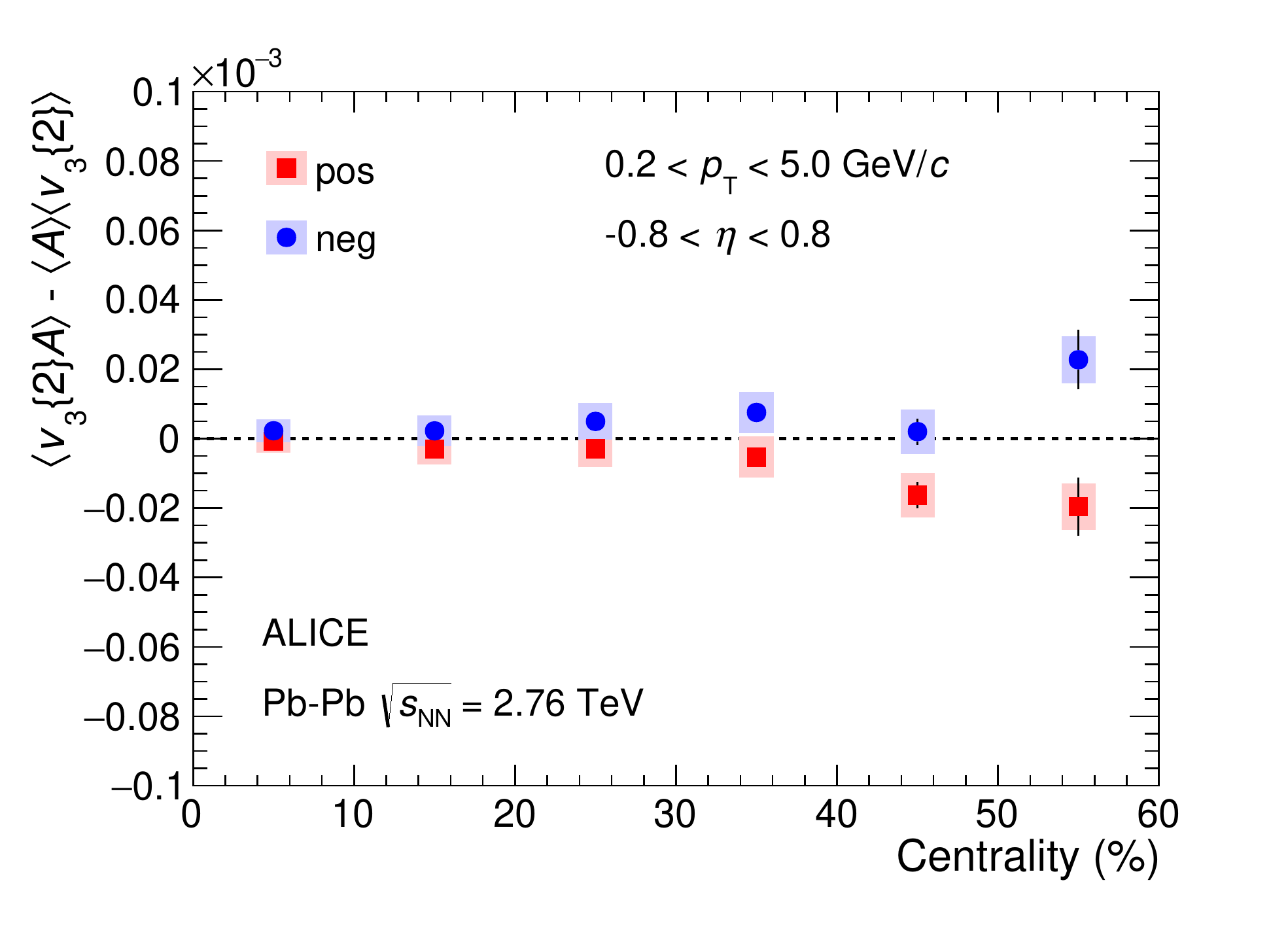}
\includegraphics[width=0.49\linewidth]{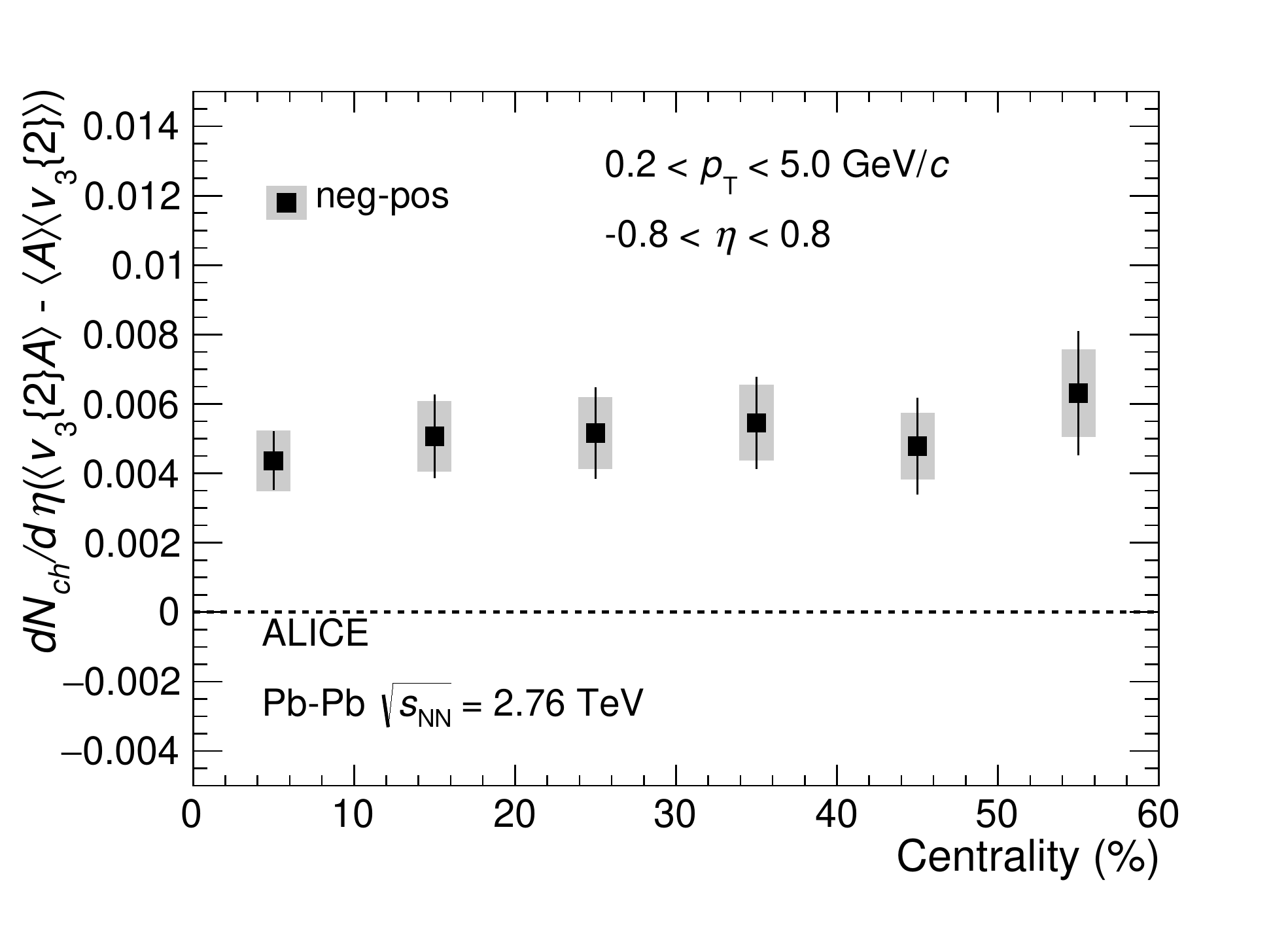}
\caption{(Color online) Three-particle correlator for the third
  harmonic (left panel) for positive (red squares) and negative (blue
  circles) particles, and the charge difference multiplied by
  $\dndeta$ (right panel).  Statistical (systematic) uncertainties are
  indicated by vertical bars (shaded boxes).}
\label{fig:c32Avscent}
\end{center}
\end{figure}

\begin{figure}[h!]
\begin{center}
\includegraphics[width=0.49\linewidth]{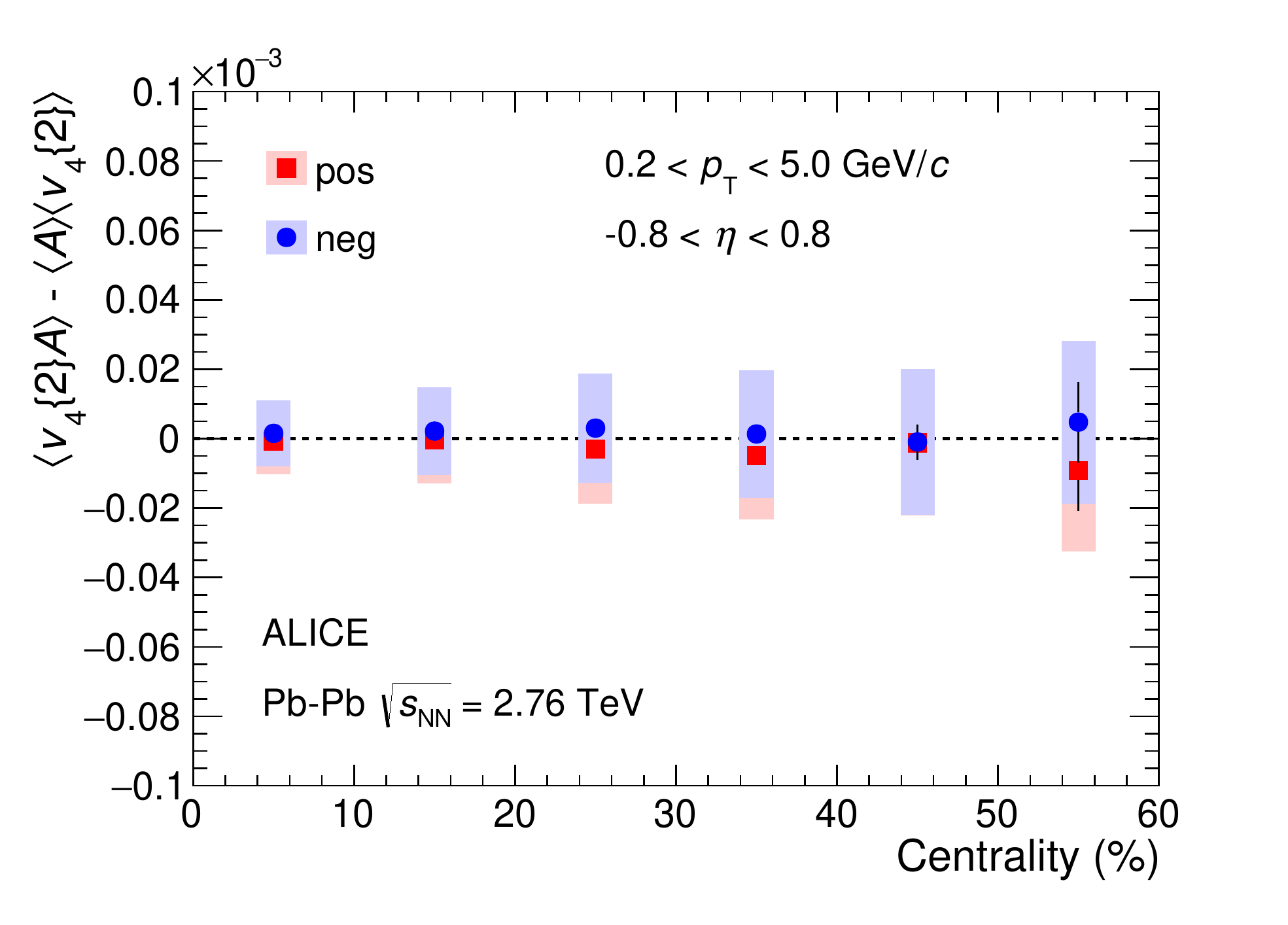}
\includegraphics[width=0.49\linewidth]{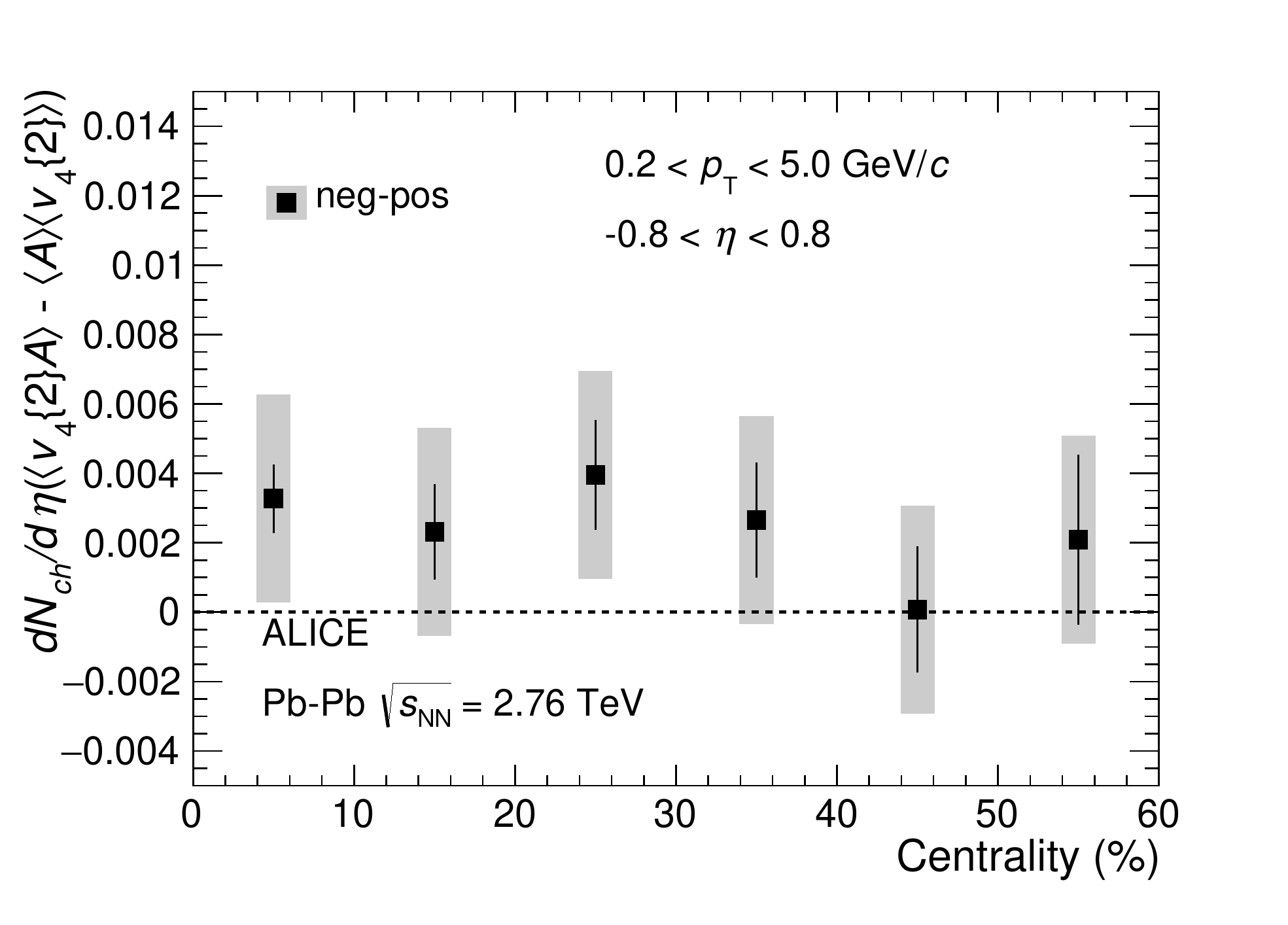}
\caption{(Color online) Three-particle correlator for the fourth
  harmonic (left panel) for positive (red squares) and negative (blue
  circles) particles, and the charge difference multiplied by
  $\dndeta$ (right panel).  Statistical (systematic) uncertainties are
  indicated by vertical bars (shaded boxes).}
\label{fig:c42Avscent}
\end{center}
\end{figure}

The three-particle correlator is studied using other harmonics as
well.  This provides important additional constraints because
P-violating effects are expected to occur with respect to the reaction
plane, therefore higher harmonics should have very little or no
correlations.  The three-particle correlator for the third harmonic is
shown in Fig.~\ref{fig:c32Avscent} and the fourth harmonic is shown in
Fig.~\ref{fig:c42Avscent}.  In both cases the left panel shows the
correlator for positive and negative charges separately, and the right
panel shows the charge difference of the correlator multiplied by
$\dndeta$.  In both of these cases, the centrality dependence of the
charge dependence is flat, in contrast to the second harmonic.  This
may suggest a different nature of the correlation.  It could also
reflect a weaker centrality dependence of $v_3$ compared to that of
elliptic flow.

\subsection{Slopes of $\Delta v_2$ vs. $A$}

\begin{figure}[h!]
\begin{center}
\includegraphics*[width=0.49\linewidth]{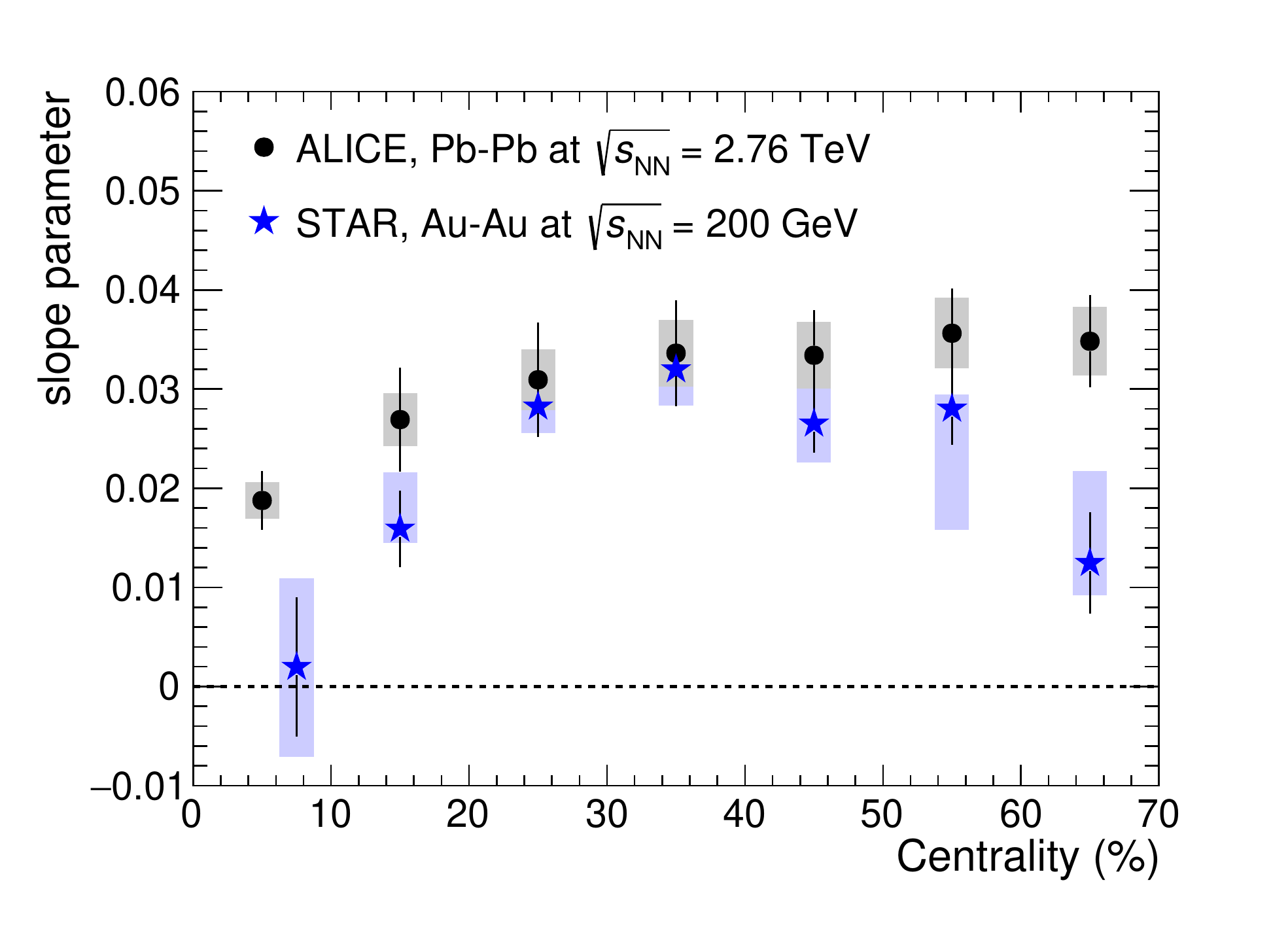}
\caption{(Color online) Slope parameter $r$ as a function of
  centrality, including points from
  STAR~\cite{Adamczyk:2015kwa}. Statistical (systematic) uncertainties
  are indicated by vertical bars (shaded boxes).  Not shown
  is the 6\% systematic normalization uncertainty
  due the MC correction for $\sigma_A^2$.  }
\label{fig:compstar}
\end{center}
\end{figure}

Figure~\ref{fig:compstar} shows a comparison between slope parameters
$r$ estimated in this analysis and from the STAR
analysis~\cite{Adamczyk:2015kwa} of \auau~collisions at
$\snn=$~200~GeV.  For the STAR data, the $v_2$ is evaluated for
charged pions with 0.15~GeV/$c$~$<\pt<$~0.5~GeV/$c$, in contrast with
the present results which are for unidentified hadrons with
0.2~GeV/$c$~$<\pt<$~5.0~GeV/$c$.  Overall, the slopes are surprisingly
similar when considering the different collision energies and
multiplicities, as well as the different kinematic acceptance (in
addition to the different $\pt$ selection, the STAR results correspond
to the pseudorapidity range $|\eta|<1.0$).
The STAR data exhibit a somewhat stronger centrality dependence than
the ALICE data.  Moreover, the STAR data exhibit a stronger centrality
dependence than predicted by the theoretical models invoking the CMW
for \auau~at 200 GeV~\cite{Burnier:2012ae}.  Additionally,
hydrodynamical models have been developed to attempt to explain the
STAR results without invoking the
CMW~\cite{Bzdak:2013yla,Hatta:2015hca}.  However, no theoretical
modeling or calculations at all, regardless of mechanism, are
available for \pbpb~collisions at $\snn=$~2.76~TeV.

\subsection{Differential correlator results as a function of $\Delta\eta$}

\begin{figure}[h!]
\begin{center}
\includegraphics*[width=0.49\linewidth]{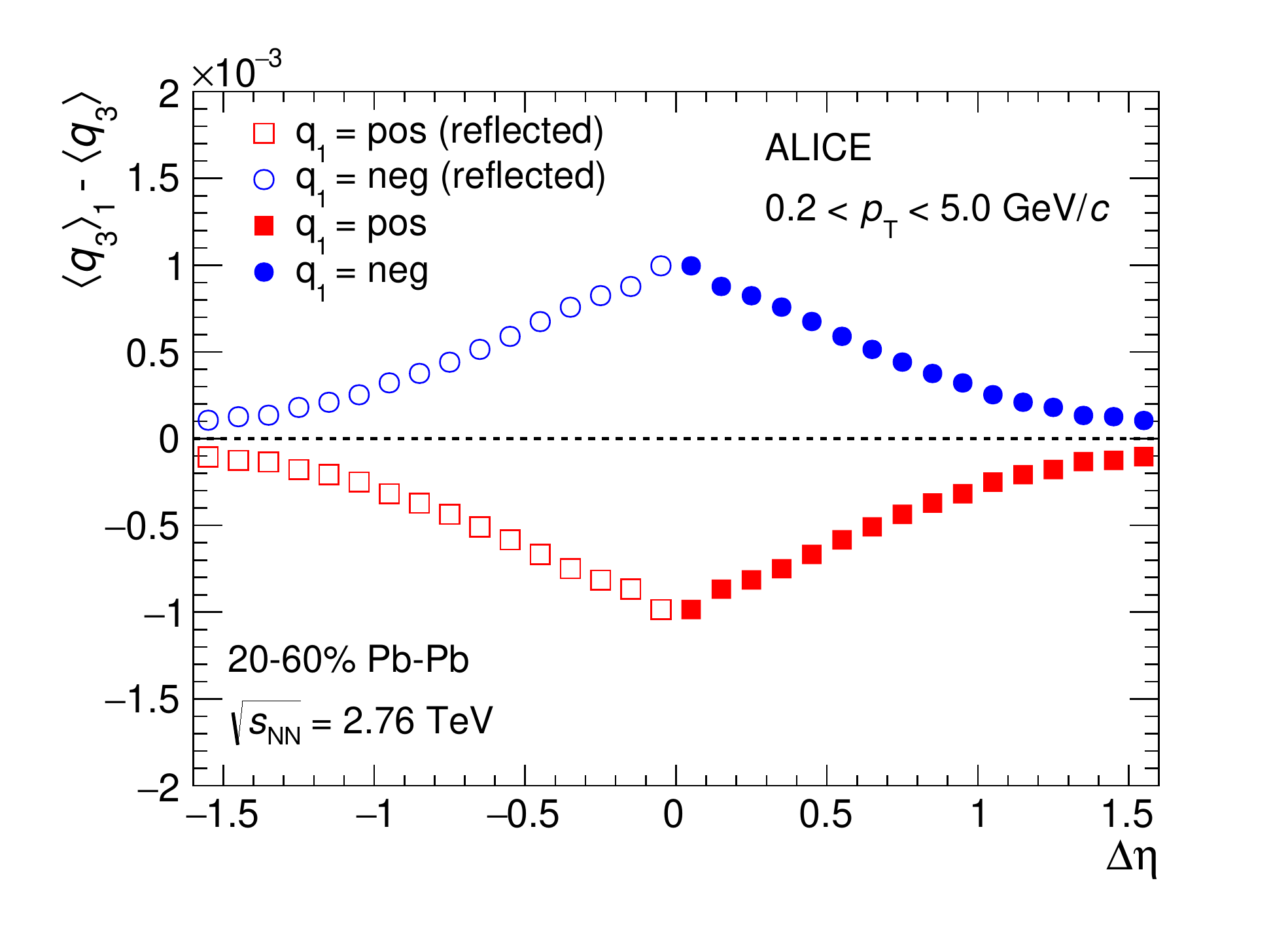}
\caption{(Color online) Correlation between the charge of the first particle $q_1$ and the
  charge of the third particle $q_3$.  Statistical uncertainties only.}
\label{fig:q3vsdeta}
\end{center}
\end{figure}

As discussed above, the definition of the three particle differential correlator includes
$\langle q_3\rangle_1$---the mean charge of the third particle when evaluated with a
selection on $q_1$.  The quantity $\langle q_3 \rangle_1 - \langle q_3 \rangle$ is shown
as a function of $\Delta\eta = \eta_1-\eta_3$ in Fig.~\ref{fig:q3vsdeta}.
The measurements are performed as a function of $|\Delta\eta|$ and shown as a function of
$\Delta\eta$ with the points reflected about $\Delta\eta=0$.  This conditional mean of
$q_3$ depends significantly on $\Delta\eta$ and has the opposite sign when $q_1$ is
flipped.  The effect is most pronounced for $\Delta\eta \approx$ 0, and the weakest when
$\Delta\eta$ is large.  When the first particle is negative, the third particle has a
slightly positive mean charge, and when the first particle is positive, the third particle
has a slightly negative mean charge.  Note that the quantity $\langle q_3 \rangle_1 -
\langle q_3 \rangle$ is proportional to the charge balance function
~\cite{Voloshin:2014gja} and as such reflects the charge correlation length.

\begin{figure}[h!]
\begin{center}
\includegraphics*[width=0.49\linewidth]{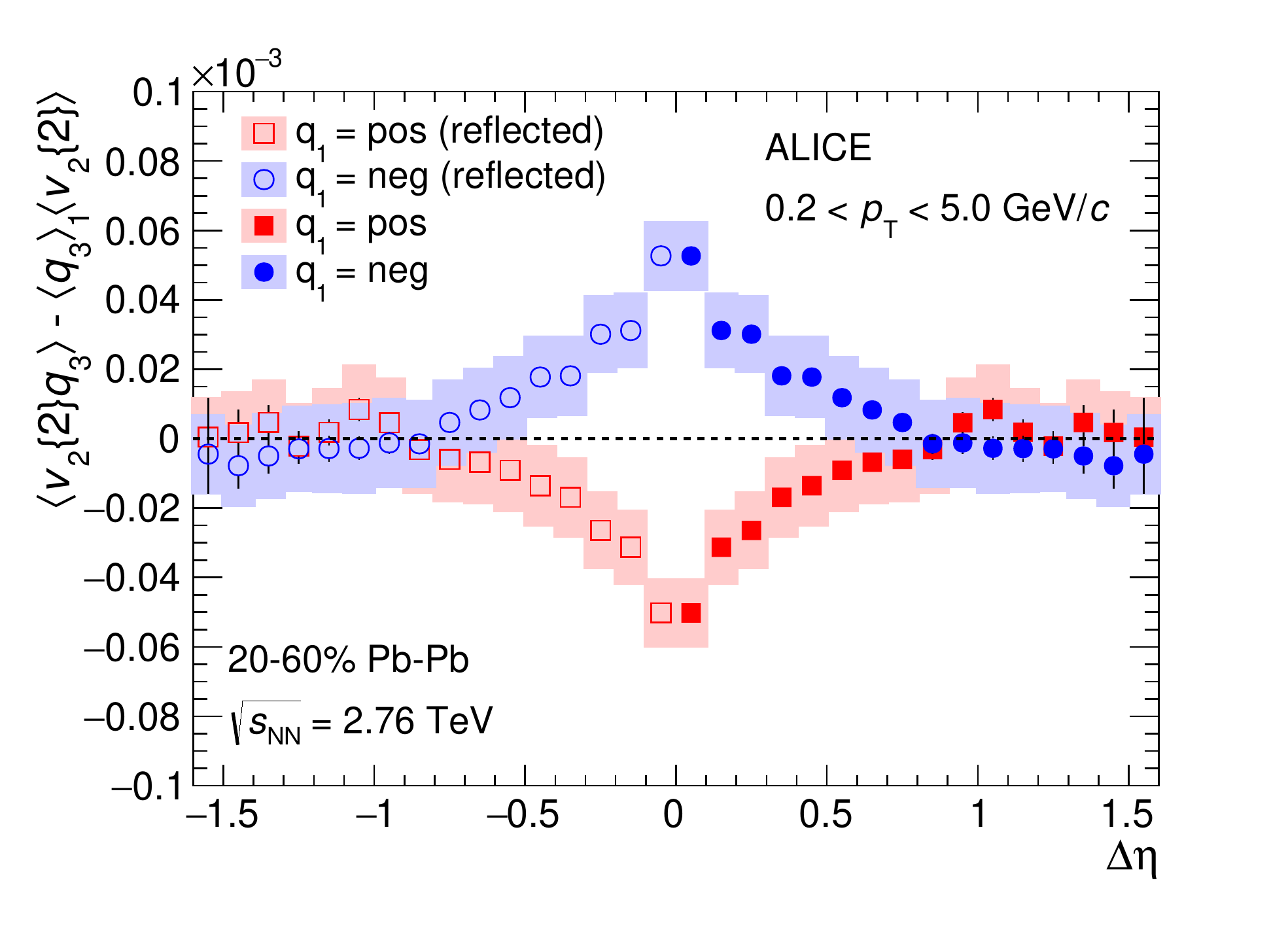}
\caption{(Color online) Three-particle correlator for the second harmonic, for positive
  (red squares) and negative (blue circles) particles.  Statistical (systematic)
  uncertainties are indicated by vertical bars (shaded boxes).}
\label{fig:c22Avsdeta}
\end{center}
\end{figure}

Figure~\ref{fig:c22Avsdeta} shows the three-particle correlator for the second harmonic as
a function of $\Delta\eta$.  The correlator exhibits a rather non-trivial dependence on
$\deta$: a peak with a ``typical hadronic width'' of about 0.5--1 units of rapidity and a
possible change of the sign at about $\deta\approx 1$ (note however these points are
consistent with zero within the systematic uncertainties).  Both of those features
qualitatively agree with possible background contribution from local charge conservation
combined with strong radial and elliptic flow~\cite{Voloshin:2014gja}. Unfortunately there
exist no predictions for this observable from the CMW.

\begin{figure}[h!]
\begin{center}
\includegraphics*[width=0.49\linewidth]{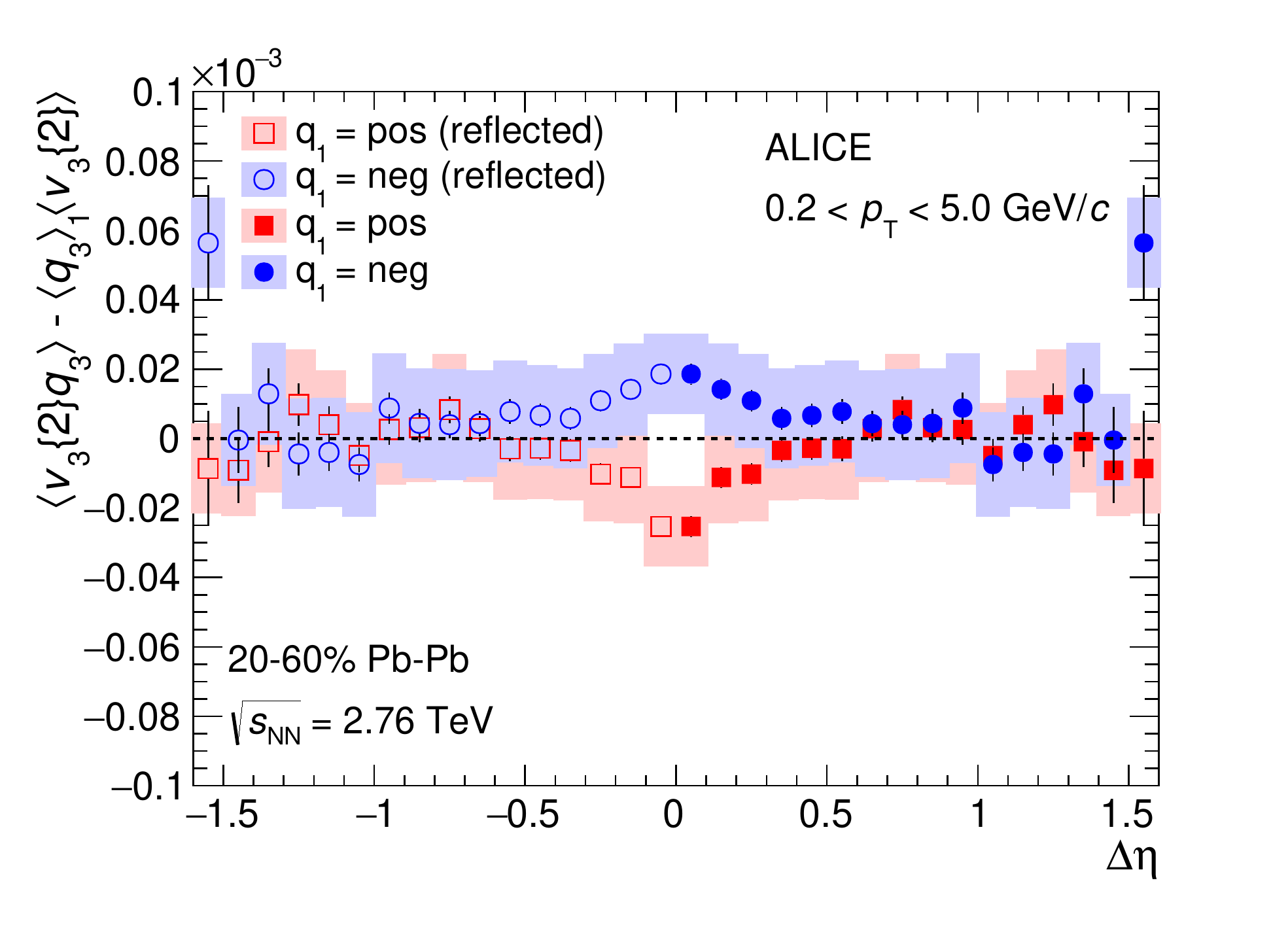}
\includegraphics*[width=0.49\linewidth]{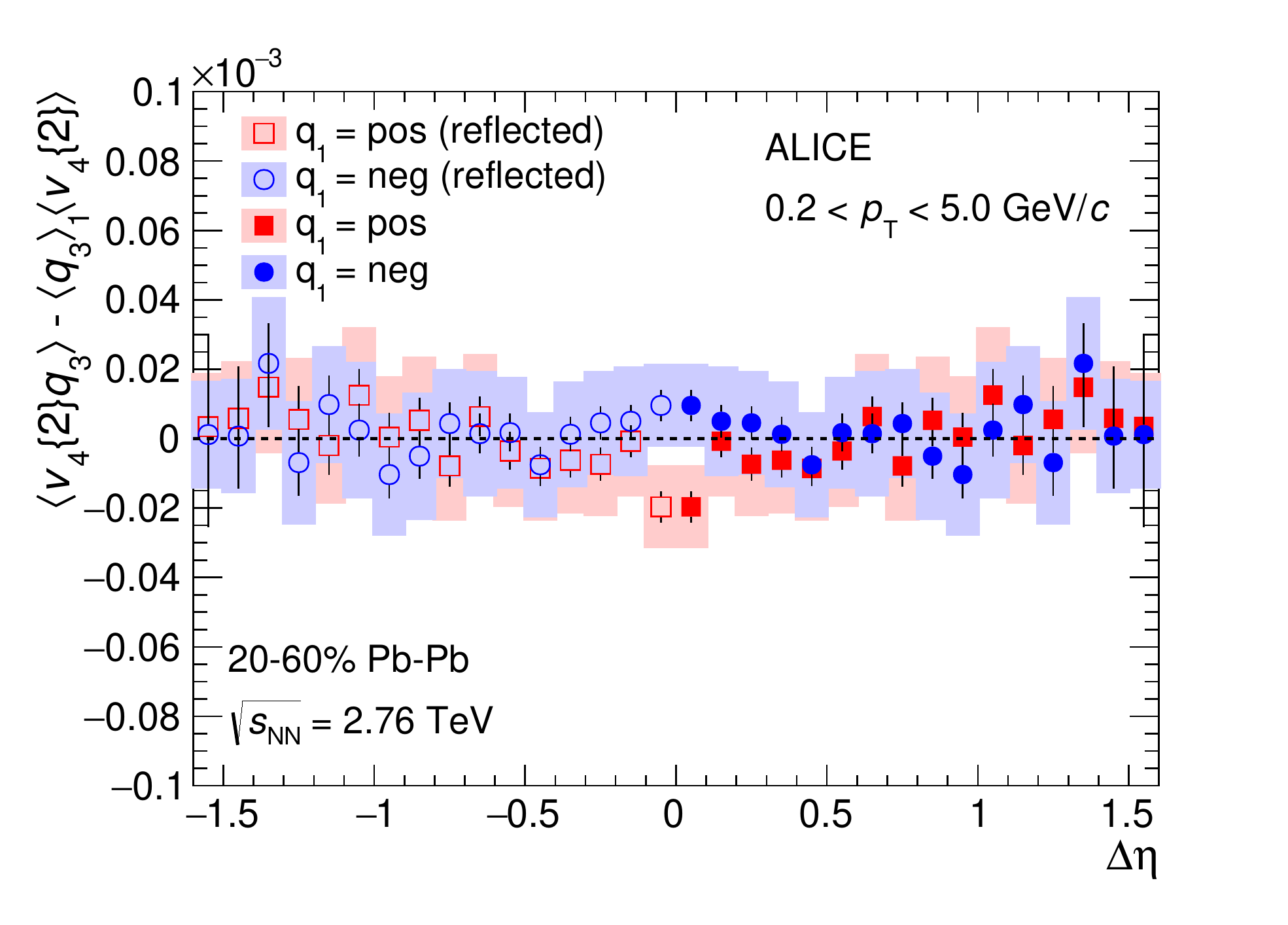}
  \caption{(Color online) Three-particle correlator for the third (left panel) and fourth
    (right panel) harmonics, for positive (red squares) and negative (blue circles)
    particles.  Statistical (systematic) uncertainties are indicated by vertical bars
    (shaded boxes).}
  \label{fig:c34deta}
  \end{center}
  \end{figure}

The three-particle correlator for the third and fourth harmonics as a function of
$\Delta\eta$ is shown in Fig.~\ref{fig:c34deta}.  The strength of the correlations is
significantly reduced, by a factor about 3 in the case of the third harmonic and at least
a factor of 5 for the fourth harmonic. The fourth harmonic correlator is consistent with
zero within errors. Neglecting flow fluctuations, the CMW expectations for higher
harmonics correlators would be zero; unfortunately there are no reliable calculations of
the effect of flow fluctuations. The (background) contribution due to the local charge
conservation should roughly scale with the magnitude of the flow~\cite{Voloshin:2014gja}
and is qualitatively consistent with the experimental results. More detailed calculations
in both scenarios, as well as more precise data, are obviously needed for a more
definitive conclusion.

\section{Summary and Outlook}

Novel three-particle correlators have been employed in an experimental search for the CMW.
Results have been shown for the second, third, and fourth harmonic for the integrated
correlator of the charge-dependent flow as a function of centrality and the differential
correlator as a function of pseudorapidity separation.
A clear dependence of the positive and negative particle anisotropic flow on the event
charge asymmetry is presented for different centralities in \pbpb~collisions.
The slopes of this dependence, determined by two different methods, are consistent and
qualitatively agree with the expectations for CMW, as well as similar to those measured by
the STAR Collaboration at the top RHIC energy. The observed nonzero signal in higher
  harmonics correlations indicates a possible strong background contribution, likely from
  LCC in combination with strong radial and anisotropic flow.
We also have presented results on the differential correlator, which is more sensitive to
the detail of the underlying physics and help to discriminate between the CMW scenario and
the background effects.  The second harmonic results show a fairly large correlation, and
the strength of the correlation strongly decreases with increasing harmonic number.
Further input from theory is needed to give detailed constraints on the magnitude and
range of background vs. CMW correlations.

LHC Run-2 will include \pbpb~collisions at $\snn=$~5.02~TeV and will offer substantially
higher integrated luminosity, which will largely improve statistical precision of these
measurements, and may also help reduce some of the systematic uncertainties.  One of the
chief benefits of increased statistical precision would be the possibility to evaluate the
three-particle correlator with identified particles.  The species of both particle 1 and 3
is of potential interest.  The different collision energy affects both the peak strength
(which increases) and the lifetime (which decreases) of the magnetic field induced in the
collision, which provides additional information.  For that reason, an analysis of this
correlator at lower collisions energies, for example at RHIC, would provide important
additional insights.

\newenvironment{acknowledgement}{\relax}{\relax}
\begin{acknowledgement}
\section*{Acknowledgements}

The ALICE Collaboration would like to thank all its engineers and technicians for their invaluable contributions to the construction of the experiment and the CERN accelerator teams for the outstanding performance of the LHC complex.
The ALICE Collaboration gratefully acknowledges the resources and support provided by all Grid centres and the Worldwide LHC Computing Grid (WLCG) collaboration.
The ALICE Collaboration acknowledges the following funding agencies for their support in building and
running the ALICE detector:
State Committee of Science,  World Federation of Scientists (WFS)
and Swiss Fonds Kidagan, Armenia;
Conselho Nacional de Desenvolvimento Cient\'{\i}fico e Tecnol\'{o}gico (CNPq), Financiadora de Estudos e Projetos (FINEP),
Funda\c{c}\~{a}o de Amparo \`{a} Pesquisa do Estado de S\~{a}o Paulo (FAPESP);
National Natural Science Foundation of China (NSFC), the Chinese Ministry of Education (CMOE)
and the Ministry of Science and Technology of China (MSTC);
Ministry of Education and Youth of the Czech Republic;
Danish Natural Science Research Council, the Carlsberg Foundation and the Danish National Research Foundation;
The European Research Council under the European Community's Seventh Framework Programme;
Helsinki Institute of Physics and the Academy of Finland;
French CNRS-IN2P3, the `Region Pays de Loire', `Region Alsace', `Region Auvergne' and CEA, France;
German Bundesministerium fur Bildung, Wissenschaft, Forschung und Technologie (BMBF) and the Helmholtz Association;
General Secretariat for Research and Technology, Ministry of Development, Greece;
National Research, Development and Innovation Office (NKFIH), Hungary;
Department of Atomic Energy and Department of Science and Technology of the Government of India;
Istituto Nazionale di Fisica Nucleare (INFN) and Centro Fermi -
Museo Storico della Fisica e Centro Studi e Ricerche ``Enrico Fermi'', Italy;
Japan Society for the Promotion of Science (JSPS) KAKENHI and MEXT, Japan;
Joint Institute for Nuclear Research, Dubna;
National Research Foundation of Korea (NRF);
Consejo Nacional de Cienca y Tecnologia (CONACYT), Direccion General de Asuntos del Personal Academico(DGAPA), M\'{e}xico, Amerique Latine Formation academique - 
European Commission~(ALFA-EC) and the EPLANET Program~(European Particle Physics Latin American Network);
Stichting voor Fundamenteel Onderzoek der Materie (FOM) and the Nederlandse Organisatie voor Wetenschappelijk Onderzoek (NWO), Netherlands;
Research Council of Norway (NFR);
National Science Centre, Poland;
Ministry of National Education/Institute for Atomic Physics and National Council of Scientific Research in Higher Education~(CNCSI-UEFISCDI), Romania;
Ministry of Education and Science of Russian Federation, Russian
Academy of Sciences, Russian Federal Agency of Atomic Energy,
Russian Federal Agency for Science and Innovations and The Russian
Foundation for Basic Research;
Ministry of Education of Slovakia;
Department of Science and Technology, South Africa;
Centro de Investigaciones Energeticas, Medioambientales y Tecnologicas (CIEMAT), E-Infrastructure shared between Europe and Latin America (EELA), 
Ministerio de Econom\'{i}a y Competitividad (MINECO) of Spain, Xunta de Galicia (Conseller\'{\i}a de Educaci\'{o}n),
Centro de Aplicaciones Tecnológicas y Desarrollo Nuclear (CEA\-DEN), Cubaenerg\'{\i}a, Cuba, and IAEA (International Atomic Energy Agency);
Swedish Research Council (VR) and Knut $\&$ Alice Wallenberg
Foundation (KAW);
Ukraine Ministry of Education and Science;
United Kingdom Science and Technology Facilities Council (STFC);
The United States Department of Energy, the United States National
Science Foundation, the State of Texas, and the State of Ohio;
Ministry of Science, Education and Sports of Croatia and  Unity through Knowledge Fund, Croatia;
Council of Scientific and Industrial Research (CSIR), New Delhi, India;
Pontificia Universidad Cat\'{o}lica del Per\'{u}.
\end{acknowledgement}

\bibliographystyle{utphys}   
\bibliography{refs}

\newpage
\appendix
\section{The ALICE Collaboration}
\label{app:collab}
\begin{flushleft} 

\bigskip 

J.~Adam$^{\rm 40}$, 
D.~Adamov\'{a}$^{\rm 84}$, 
M.M.~Aggarwal$^{\rm 88}$, 
G.~Aglieri Rinella$^{\rm 36}$, 
M.~Agnello$^{\rm 110}$, 
N.~Agrawal$^{\rm 48}$, 
Z.~Ahammed$^{\rm 132}$, 
S.U.~Ahn$^{\rm 68}$, 
S.~Aiola$^{\rm 136}$, 
A.~Akindinov$^{\rm 58}$
% ...

\bigskip

\bigskip 

\textbf{\Large Affiliation Notes}

\bigskip 

% ...
$^{\rm IV}$ Also at: M.V. Lomonosov Moscow State University, D.V. Skobeltsyn Institute of Nuclear, Physics, Moscow, Russia

\bigskip

\bigskip 

\textbf{\Large Collaboration Institutes}

\bigskip 

% ...
$^{36}$ European Organization for Nuclear Research (CERN), Geneva, Switzerland\\
% ...
$^{40}$ Faculty of Nuclear Sciences and Physical Engineering, Czech Technical University in Prague, Prague, Czech Republic\\
% ...
$^{48}$ Indian Institute of Technology Bombay (IIT), Mumbai, India\\
% ...
$^{58}$ Institute for Theoretical and Experimental Physics, Moscow, Russia\\
% ...
$^{68}$ Korea Institute of Science and Technology Information, Daejeon, South Korea\\
% ...
$^{84}$ Nuclear Physics Institute, Academy of Sciences of the Czech Republic, \v{R}e\v{z} u Prahy, Czech Republic\\
% ...
$^{88}$ Physics Department, Panjab University, Chandigarh, India\\
% ...
$^{110}$ Sezione INFN, Turin, Italy\\
% ...
$^{132}$ Variable Energy Cyclotron Centre, Kolkata, India\\
% ...
$^{136}$ Yale University, New Haven, Connecticut, United States\\
% ...

\bigskip 

\end{flushleft} 

\end{document}